\documentclass[]{aa}  
\usepackage{natbib} 
\usepackage{graphicx}
\usepackage{txfonts}
\def\ax{AX~J1910.7+0917}
\def\inte{{\em INTEGRAL}}
\def\xmm{{\em XMM-Newton}}

\def\chan{{\em Chandra}}
\def\asca{{\em ASCA}}

\def\swift{{\em Swift}}
\def\rosat{{\em ROSAT}}
\def\einst{{\em Einstein}}

\begin{document}
   \title{ \ax\ and three newly discovered INTEGRAL sources}

\author{L. Pavan   
	\inst{1,2}
	\and
	E. Bozzo 
	\inst{1,2}
	\and 
	C. Ferrigno 
	\inst{1,2}
	\and 
	C. Ricci 
	\inst{1,2}
	\and	
	A. Manousakis 
	\inst{1,2}
	\and 
        R. Walter
	\inst{1,2}
	\and	
	L. Stella
	\inst 3
}

\authorrunning{L. Pavan et al.}
  \titlerunning{\ax\ and three newly discovered INTEGRAL sources}
  \offprints{L. Pavan}

\institute{ISDC, INTEGRAL Science Data Centre,  Universit\'e de Gen\`eve, Chemin d'Ecogia 16, CH-1290 Versoix, Switzerland\\
	\email{lucia.pavan@unige.ch}
        \and Observatoire de Gen\`eve, Universit\'e de Gen\`eve, Chemin des Maillettes 51, CH-1290 Sauverny, Switzerland
         \and
          INAF - Osservatorio Astronomico di Roma, Via Frascati 33, 00044 Rome, Italy. 
         }

 \abstract{} {We take advantage of the high sensitivity of the IBIS/ISGRI telescope and the improvements in the 
 data analysis software to investigate the nature of the still poorly known X-ray source  
 \ax,\ and search for closeby previously undetected objects.}
 {We analyze all publicly available \inte\ data of \ax,\ together with a number of archival observations 
 that were carried out in the direction of the source with \chan,\ \xmm,\ and \asca.\ In the IBIS/ISGRI field-of-view  
 around \ax,\ we discovered three new sources: IGR\,J19173+0747, IGR\,J19294+1327 and 
 IGR\,J19149+1036; the latter is positionally coincident with the \einst\ source 2E\,1912.5+1031.  
 For the first two sources, we also report the results of follow-up observations carried out with \swift\,/XRT.}
 {\ax\ features a clear variability in the X-rays. Its spectrum can be well described with an  
 absorbed ($N_{\rm H}$$\sim$6$\times$10$^{22}$~cm$^{-2}$) power-law ($\Gamma$$\simeq$1.5) model plus 
 an iron line at $6.4$~keV. We also obtained a refined position and report on possible infrared counterparts.} 
 {The present data do not allow for a unique classification of the sources. 
 Based on the property of its X-ray emission and the analysis of a likely infrared counterpart, we investigate different 
 possibilities for the nature of \ax.\ }
   

   \date{Received 2010 August 8; accepted 2010 November 27}

   \maketitle

\section{Introduction}
\label{sec:intro}

The wide field-of-view of the IBIS/ISGRI telescope 
\citep[FOV, 19$^{\circ}$$\times$19$^{\circ}$;][]{ubertini03} onboard 
\inte\ \citep{winkler03} and its unprecedented sensitivity in the hard X-ray 
domain (17-100~keV) have made this instrument 
particularly successful in the past few years in revealing new high-energy 
sources.  The latest available IBIS/ISGRI catalog contains more than 
700 objects \citep{bird09} and an increasing number of sources are being 
discovered thanks to on-going \inte\ surveys of the high-energy 
sky. Since 2003, the relatively large exposure time 
available ($\sim$1-6~Ms) permitted to achieve in the galactic plane
 a limiting sensitivity of $\lesssim$1~mCrab in the 17-100~keV energy range 
and a point-source location accuracy of $\simeq$2-3 arcmin. 
Besides the increasing amount of observing time, the identification 
of a number of new hard X-ray sources has benefited from improvements in the data 
analysis software.   

In this paper, we take advantage of the new version of the \inte\ OSA 
software released by the ISDC \citep{courvoisier03} 
to investigate the nature of the still poorly known source \ax,\ and 
search for nearby previously undetected faint X-ray sources in the 
field-of-view (FOV) around this source. 
In Sect.~\ref{sec:ax} we summarize all previous observations of \ax\ and
 the analysis of all the publicly available 
\inte\ data.\ We also carried out an analysis of all  
archival \chan,\ \xmm,\ and \asca,\ observations that included  
\ax\ in the instruments FOV. 
The analysis of the \inte\ data also led to the discovery of three new hard X-ray 
sources in the IBIS/ISGRI FOV around \ax, independently detected through data analysis also with the    
{\sc bat\_imager} software (A.~Segreto, private communication).
 For two of these sources, we obtained 
follow-up observations with \swift\,/XRT and present these results 
in Sect.~\ref{sec:newsources}. Our discussion and conclusion 
are summarized in Sect.~\ref{sec:discussion}.

\section{ \ax }
\label{sec:ax} 

\ax\ is a relatively faint and poorly known X-ray source discovered with \asca\ during 
the survey of the Galactic plane. The best-determined position so far is at 
$\alpha_{\rm J2000}$=19$^{\rm h}$10$^{\rm m}$47$\fs$00 and 
$\delta_{\rm J2000}$=09${\degr}$17$\arcmin$06$\farcs$0, with an associated error of 57.6$\farcs$ 
\citep[90\% c.l.,][]{sugizaki01}. 
The \asca\ spectrum could be fitted with an absorbed power-law model 
($\chi^2_{\rm red}$/d.o.f.=0.72/10). The measured photon index was 
$\Gamma$=1.1$^{+0.5}_{-0.4}$, for an absorption column density 
of $N_{\rm H}$=(2.6$^{+1.4}_{-1.0}$)$\times$10$^{22}$~cm$^{-2}$. 
\begin{figure*}[ht!]
  \centering
  \includegraphics[width=17cm]{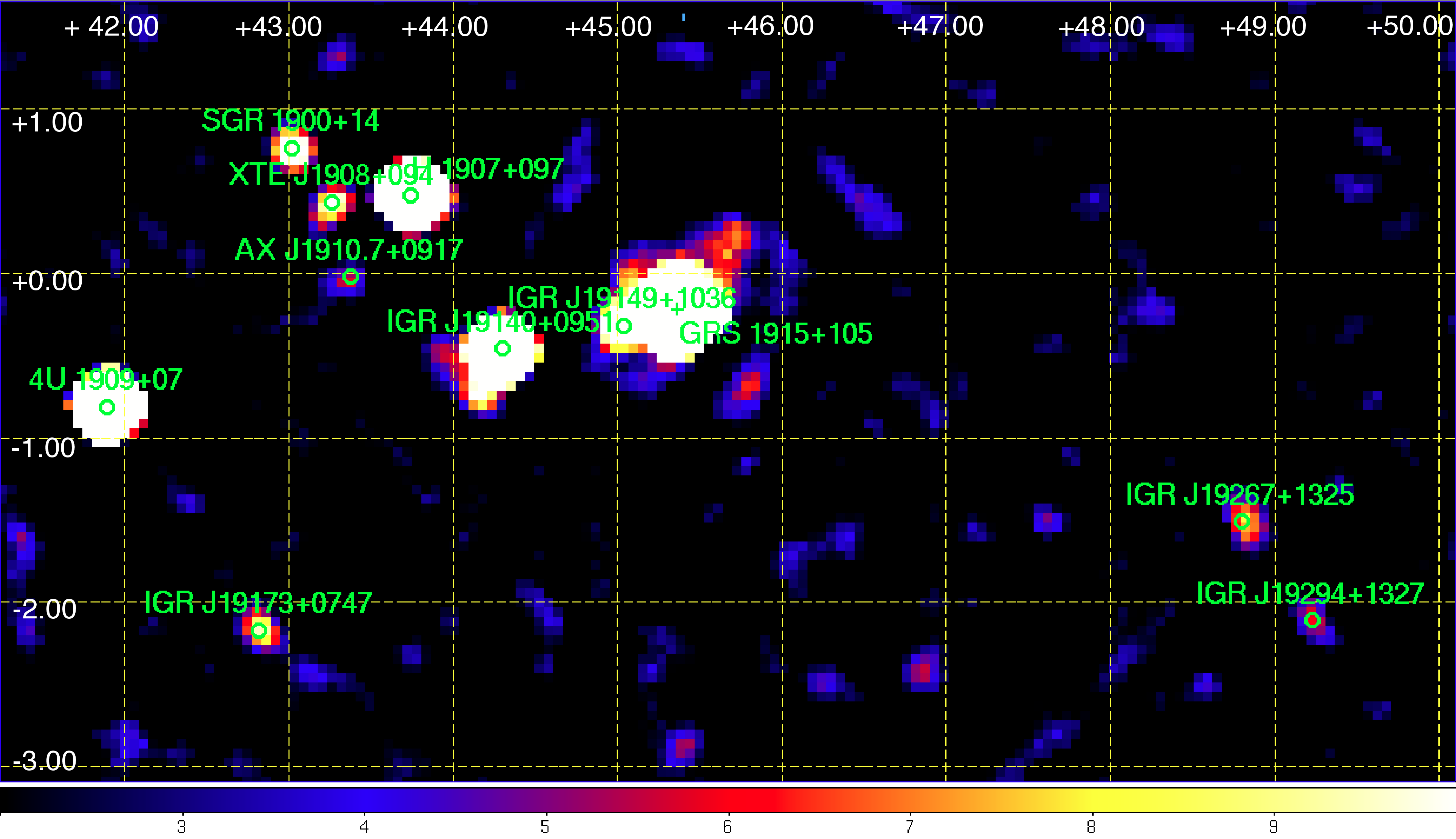}
  \caption{IBIS/ISGRI mosaic around \ax\ (17-80~keV, significance map).\ We also 
show the newly discovered nearby sources 
(see Sect.~\ref{sec:newsources} and the electronic version   
of the paper for the colored picture).
 The dashed grids denote galactic coordinates.}
  \label{fig:mosaic}
\end{figure*}
The estimated 0.7-10~keV X-ray flux was 2.4$\times$10$^{-12}$~erg/cm$^{-2}$/s. 
\citet{sugizaki01} associated \ax\ to the \einst\ source 2E\,1908.3+0911,  
which is 
%located at $\alpha_{\rm J2000}$=19$^{\rm h}$08$^{\rm m}$22$\fs$00 and 
%$\delta_{\rm J2000}$=09${\degr}$11$\arcmin$13$\farcs$0 (90\% c.l. error 45$\farcs$) and 
characterized by 
an averaged X-ray flux of (6.9$\pm$1.4)$\times$10$^{-13}$~erg/cm$^{2}$/s \citep[0.2-3.5~keV,][]{hertz88}. 
The source was also detected with \inte,\ and classified by \citet{bird09} as a likely variable source.  
 The source maximum detection significance in the IBIS/ISGRI data is 8.3~$\sigma$ (18-60~keV), 
and the corresponding time-average flux is 0.4$\pm$0.1~mCrab (0.6$\pm$0.1~mCrab) in the 20-40~keV 
(40-100~keV) energy band (corresponding to 3.0$\times$10$^{-12}$~erg/cm$^2$/s and 
5.7$\times$10$^{-12}$~erg/cm$^2$/s, respectively).  
No other detections and counterparts in different energy band 
have been reported so far.

\subsection{Data analysis and results}
\label{sec:data}

We analyzed all archival \inte,\ \asca,\ \xmm,\ and \chan\  
observations that included the position of \ax\ in their FOV. 
This is the first in-depth study of the emission from this source up to hard X-ray energies. 
In Sect.~\ref{sec:integral} we report the details of the \inte\ observations.
A log of all other observations is 
given in Table~\ref{tab:log}. Throughout the paper, all uncertainties 
are given at 90\% c.l. (unless otherwise stated).

\begin{table*}
\caption{ Observation log of \ax.\ } 
\label{tab:log}
\centering
\begin{tabular}{@{}ccccccccc@{}}
\hline\hline
\noalign{\smallskip}
OBS ID & Instr & Date\tablefootmark{a} & Exp\tablefootmark{b} & $N_{\rm H}$ & $\Gamma$ &  $F_{\rm obs}$\tablefootmark{c} & $\chi2_{\rm red}$/d.o.f. \\
        &    &           &  (ks)   &  (10$^{22}$~cm$^{-2}$) & & (erg/cm$^{2}$/s) & (C-stat/d.o.f.) \\
\noalign{\smallskip}
\hline
\noalign{\smallskip}
\multicolumn{2}{c}{\asca} \\
\noalign{\smallskip}
50005000 & GIS2+GIS3 & 1993-04-24 & 95 & 4.8$^{-1.5}_{+1.9}$ &  1.4$\pm$0.5 & 8.2$_{-2.7}^{+0.7}$ & 0.4/53\\
\noalign{\smallskip}
13000000\tablefootmark{d} & GIS2 & 1993-05-03 & 2.3 & 4.8 (fixed) &  1.4 (fixed) &   $<$4.2 & --- \\
                 & GIS3 & 1993-05-03 & 2.3 & 4.8 (fixed) &  1.4 (fixed) &   $<$4.9 & --- \\
\noalign{\smallskip}
13000010\tablefootmark{d} & GIS2 & 1993-05-03 & 2.3 & 4.8 (fixed) &  1.4 (fixed) &  $<$1.1 & --- \\
         & GIS3 & 1993-05-03 & 2.3 & 4.8 (fixed) &  1.4 (fixed) &  $<$1.6 & --- \\
\noalign{\smallskip}
13000030\tablefootmark{d} & GIS3 & 1993-05-03 & 1.4 & 4.8 (fixed) &  1.4 (fixed) &   $<$2.1 & --- \\
         & GIS3 & 1993-05-03 & 1.4 & 4.8 (fixed) &  1.4 (fixed) &   $<$3.1 & --- \\
\noalign{\smallskip}
50005010\tablefootmark{d} & GIS2 & 1993-10-16 & 12 & 4.8 (fixed) &  1.4 (fixed) &   $<$0.6 & --- \\
         & GIS3 & 1993-10-16 & 12 & 4.8 (fixed) &  1.4 (fixed) &   $<$0.6 & ---  \\
\noalign{\smallskip}
50005020\tablefootmark{d} & GIS2 & 1993-10-17 & 20 & 4.8 (fixed) &  1.4 (fixed) &   $<$1.1 & --- \\
         & GIS3 & 1993-10-17 & 20 & 4.8 (fixed) &  1.4 (fixed) &   $<$1.4 & ---  \\
\noalign{\smallskip}
10020000 & GIS2+GIS3 & 1993-11-03 & 26 & 6.3$_{-1.4}^{+1.6}$ &  2.3$\pm$0.5 & 4.9$^{-4.3}_{+0.4}$ & (146.1/157) \\
\noalign{\smallskip}
10020010\tablefootmark{d} & GIS2 & 1993-11-03 & 19 & 4.8 (fixed) &  1.4 (fixed) &   $<$2.1 & --- \\
         & GIS3 & 1993-11-03 & 19 & 4.8 (fixed) &  1.4 (fixed) &   $<$2.3 & --- \\
\noalign{\smallskip}
57005050 & GIS2+GIS3 & 1999-04-27 & 23 & 5.0$_{-2.4}^{+3.4}$ & 1.6$\pm$0.8 & 2.5$_{-1.4}^{+0.2}$ & (70.7/59) \\
\noalign{\smallskip}
\multicolumn{2}{c}{\xmm}\\
\noalign{\smallskip}
0084100401 & Epic-pn & 2004-04-03  & 14.0 & 6.3$^{+0.5}_{-0.4}$ &  1.4$\pm$0.1 & 17.1$^{+1.0}_{-2.1}$ & 1.1/147 \\
\noalign{\smallskip}
0084100501\tablefootmark{e} & Epic-pn & 2004-04-05 & 14.7 & 5.0$\pm$0.3 & 1.28$\pm$0.08 & 24.3$_{-1.7}^{+1.2}$ & 0.9/168 \\
\noalign{\smallskip}
\multicolumn{2}{c}{ \chan\ }\\
\noalign{\smallskip}
9615\tablefootmark{f} & ACIS-S & 2008-05-31 & 1.7 & --- &  --- &   $<$0.4 & --- \\
\noalign{\smallskip}
\hline
\end{tabular}
\tablefoot{
\tablefoottext{a}{Format is YYYY-MM-DD; }
\tablefoottext{b}{Exp indicates the total exposure time of each observation; }
\tablefoottext{c}{Observed flux in the 1-10\,keV energy band in units of 10$^{-12}$; }
\tablefoottext{d}{90\% c.l. upper limit; }
\tablefoottext{e}{This fit includes also a Gaussian line at $\sim$6.4~keV, see text for details; }
\tablefoottext{f}{68\%~c.l. upper limit.}
}
\end{table*}

\subsubsection{INTEGRAL}
\label{sec:integral}

\inte\ observations are commonly divided into 
$\sim$2-3~ks short pointings called ``science windows'' (SCWs).  
We considered all publicly available SCWs for the IBIS/ISGRI 
\citep[17-80~keV,][]{lebrun03} and for the two JEM-X telescopes 
\citep[3-23~keV,][]{lund03} that were performed in the direction of \ax\ 
from 2003 March 6 to 2009 April 15 (see Sect.~\ref{sec:ax}).
This permitted us to achieve an effective exposure time 
on the source of 4.8$\times$10$^2$~ks and 2.7$\times$10$^3$~ks 
for JEM-X and ISGRI respectively. 
All data, here and below, were analyzed with version 9.0 of the OSA 
software \citep{courvoisier03}. 

The source was not detected in the JEM-X1 and JEM-X2 mosaics, and  
we derived an upper limit on the source X-ray flux by using the tool 
{\sc mosaic\_spec}. From the JEM-X1 mosaic we derived a 3~$\sigma$ upper limit 
of 1.0$\times$10$^{-11}$~erg/cm$^2$/s, 5.8$\times$10$^{-12}$~erg/cm$^2$/s, and 
9$\times$10$^{-12}$~erg/cm$^2$/s respectively in the energy bands 3-7~keV, 7-11 keV, 
and 11-19 keV. The corresponding 3~$\sigma$ upper limits derived from the 
JEM-X2 mosaic were 1.7$\times$10$^{-11}$~erg/cm$^2$/s, 1.2$\times$10$^{-11}$~erg/cm$^2$/s, 
and 1.8$\times$10$^{-11}$~erg/cm$^2$/s. These are compatible with the 
measured averaged \asca\ flux reported in Sect.~\ref{sec:intro}.\\
We produced an IBIS/ISGRI mosaic of the region around \ax, using the energy band 17-80~keV
to maximize the S/N. In this energy band \ax\ is detected with a significance of 5.8$\sigma$. 
 A close view of the ISGRI mosaic is shown in Fig.~\ref{fig:mosaic}. 
We checked that the best-fit position of the source obtained from the
mosaic is compatible with both its \asca\ and \xmm\ refined position (see Sect.~\ref{sec:xmm}).\\
We derived a count rate of 0.09$\pm$0.02 cts/s, corresponding to
 a flux of 0.31$\pm$0.05~mCrab. 
This flux is higher than any contamination expected in the same energy band 
from the nearby supernova remnant \citep[SNR, see][and Sect.~\ref{sec:xmm}]{miceli06}. \\
We also performed a spectral analysis and extracted its lightcurve 
in the two energy bands 20-40~keV and 40-80 keV.  
The averaged IBIS/ISGRI spectrum of the source is shown in Fig.~\ref{fig:unfolded} and 
discussed in Sect.~\ref{sec:xmm}. 
%The estimated 20-40~keV X-ray 
%flux is 2.1$\times$10$^{-12}$~erg/cm$^{2}$/s. 
We barycenter-correct the photon arrival times in the source lightcurves 
with the OSA9 tool {\sc barycent}. No evidence 
for a coherent periodicity could be found. This is also  
confirmed by the analysis of the data from \swift\,/BAT (Cusumano, private 
communication), which operates in a similar energy band to that of IBIS/ISGRI.  

\subsubsection{XMM-Newton}
\label{sec:xmm}

\ax\ was serendipitously\footnote{\ax\ is named as 2XMM\,J191043.4+091629 in the \xmm\ 
serendipitous source catalogue \citep{watson09}.} observed in two \xmm\ 
\citep{jansen01} observations performed 
in 2004 April in the direction of the nearby SNR W49B \citep{miceli06}. 
We processed \xmm\ observation data files (ODFs) to produce calibrated
event lists using the {\sc epproc} and {\sc
emproc} tasks (Science Analysis System, SAS, v.10.0) 
for the Epic-pn and the MOS cameras. The event files of the two observations 
were filtered to exclude high background time intervals following the 
reccomandations of the SAS online analysis threads\footnote{http://xmm.esac.esa.int/sas/documentation/threads/.}.  
We excluded from further analysis time intervals in observation 0084100401 (0084100501) during which the count-rate 
of the entire detector FOV in the 10-12 keV energy band was higher than 0.2 (0.15) cnts/s for the Epic-MOS 
and 0.35 (0.4) for the Epic-pn. We also carefully checked that none of these rises in the total field count-rate 
was due to a flare from \ax.

\begin{figure}
  \resizebox{\hsize}{!}{\includegraphics{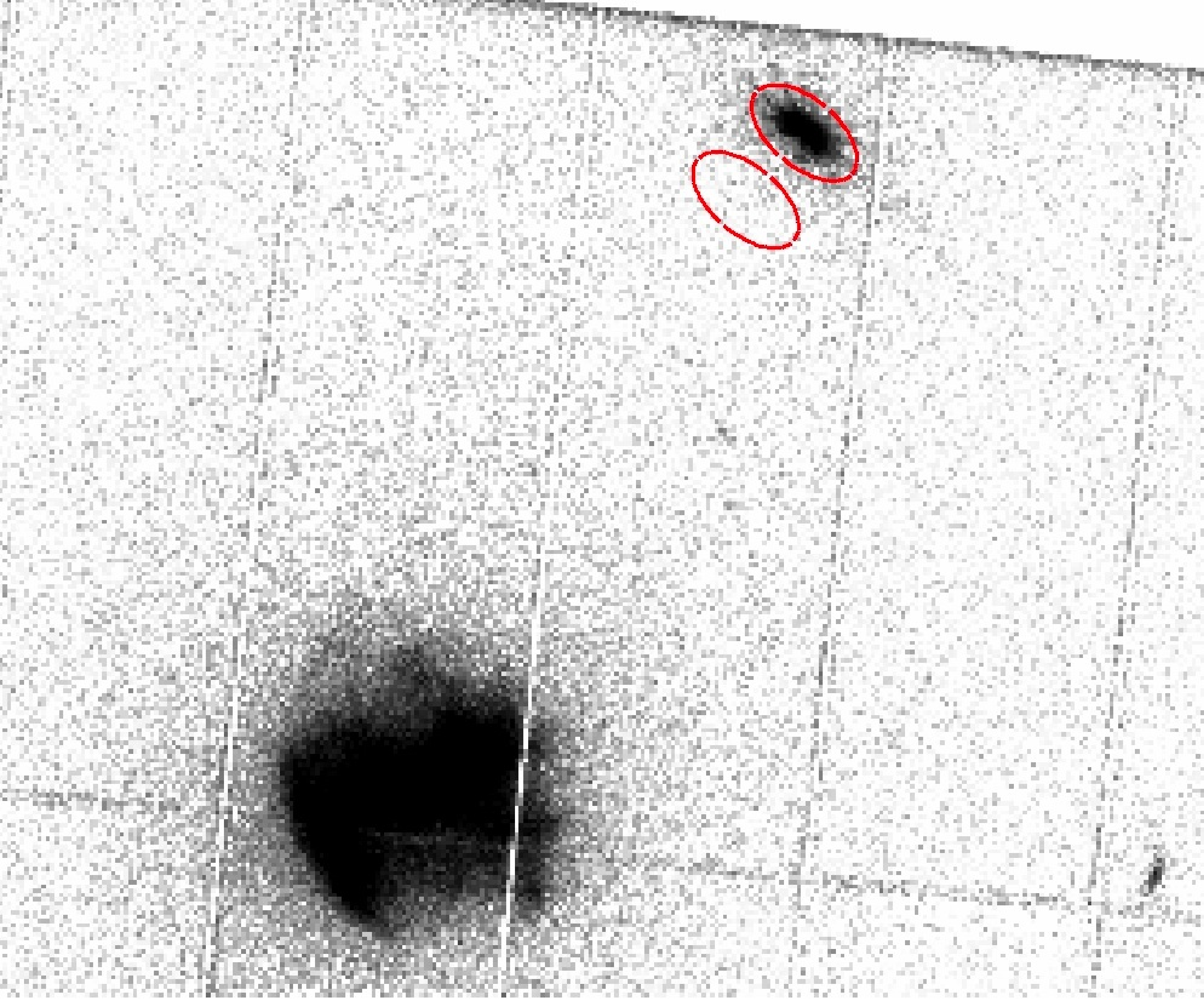}}
  \caption{Image of \ax\ as observed by the Epic-pn detector (0.5-12.0~keV). 
    The extraction region for source and background are also shown. The bright object 
    in the center is the SNR W49B.}
  \label{fig:epicpnimage} 
\end{figure}

In both observations 0084100401 and 0084100501, the three Epic cameras were 
operated in full frame and the source \ax\ was 
located at the rim of their FOV. In the Epic-MOS1 \ax\ lied on the 
border between two CCDs, and thus we excluded these data from  
the analysis. The total effective exposure time is 
of 14.0~ks (18.1~ks) for the Epic-pn (Epic-MOS2) in the 
observation 0084100401 and 14.7~ks (18.0~ks)  
in the observation 0084100501. In order to maximize S/N, source 
lightcurves and spectra were extracted by using an elliptical 
region (see Fig.~\ref{fig:epicpnimage}). 
As we discuss also below in more detail, the elongated shape of the source 
is due to the off-axis point spread function of the \xmm\ telescope 
for sources close to the border of the FOV\footnote{See also 
http://xmm.esa.int/external/xmm\_user\_support/$\!$ documentation/uhb/node18.html}.  
Background lightcurves and spectra were extracted in the 
closest source-free region to \ax. We checked that none of the results reported in this paper 
 changed significantly by using different (reasonable) source and background extraction regions.
We corrected all lightcurves for vignetting, bad pixels, point-spread-function (PSF) losses, and dead time with the SAS {\sc epiclccorr} 
tool.  All EPIC images and spectra were corrected for 
out of time events, according to the instructions provided by the SAS online analysis threads.
Epic-pn and Epic-MOS spectra were rebinned before fitting
so as to have at least 25 counts per bin and, at the same time, prevent oversampling the
energy resolution by more than a factor of three. 
Given the relatively short exposure time 
and low X-ray flux of the source, we report below 
only results from the Epic-pn analysis and checked that the Epic-MOS2 
data would give compatible results in all cases. 

\begin{figure}
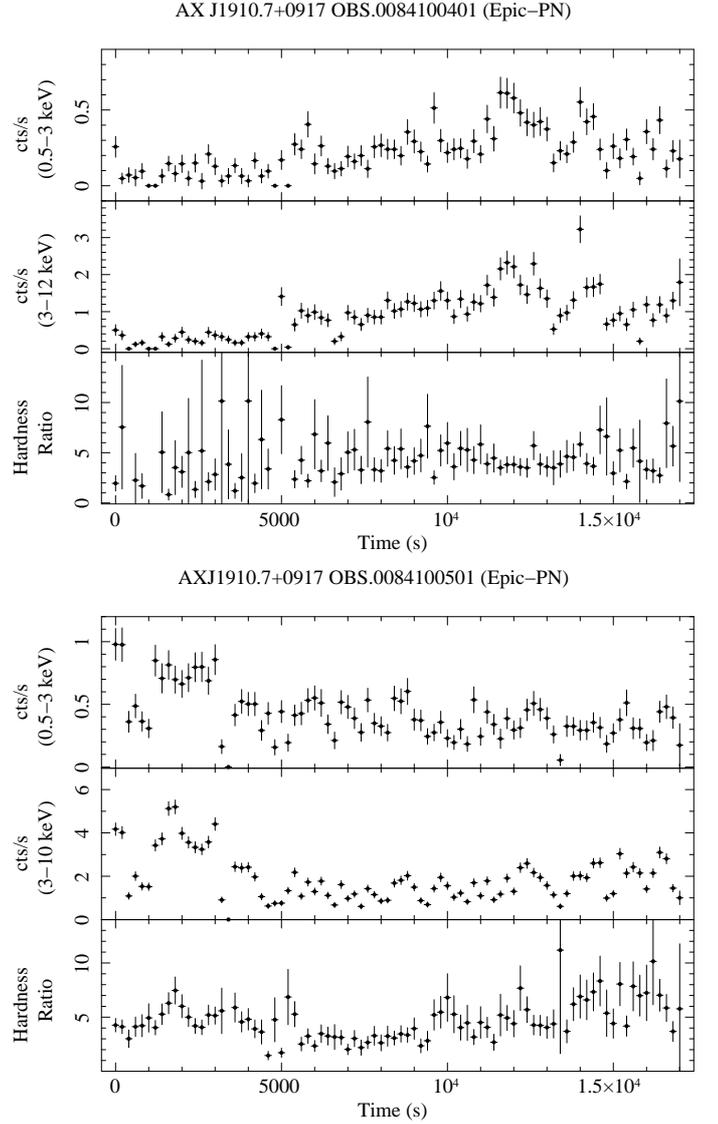

  \resizebox{\hsize}{!}{\includegraphics[angle=-90]{xmmlcurve_401.ps}}
  \resizebox{\hsize}{!}{\includegraphics[angle=-90]{xmmlcurve_501.ps}}
  \caption{ \xmm\ Epic-pn background-subtracted lightcurves of \ax\ extracted 
    in the two energy bands 0.5-3~keV and 3-12~keV for the two observations 
    0084100401 (top) and 0084100501 (bottom). The hardness ratio is reported 
    in the bottom panel of each figure. The time bin is 200 s.} 
  \label{fig:xmmlcurve}
\end{figure}

In Fig.~\ref{fig:xmmlcurve} we report the Epic-pn lightcurves of the 
source in 0.5-3~keV and 3-12.0~keV the energy bands, extracted from the 
two \xmm\ observations. The hardness ratio, defined as the ratio of the 
count rate in the hard (3-12~keV) to soft (0.5-3~keV) energy band versus time, 
is also shown. A pronounced variability on timescales of hundreds 
of seconds is clearly visible from these lightcurves, but only marginal variations 
in the hardness ratio were measured (the most prominent feature is the increase of a factor 
$\sim2$ at the beginning of observation 0084100501, see Fig.~\ref{fig:xmmlcurve}).
From the 0.5-12~keV Epic-pn lightcurves in observation 0084100401 (0084100501) 
we estimated a minimum source count rate of $0.25\pm0.03$ ($0.6\pm0.2$) cts/s 
and a maximum source count rate of $5.6\pm 0.5$ ($6.2\pm0.5$) cts/s. 

\begin{figure}
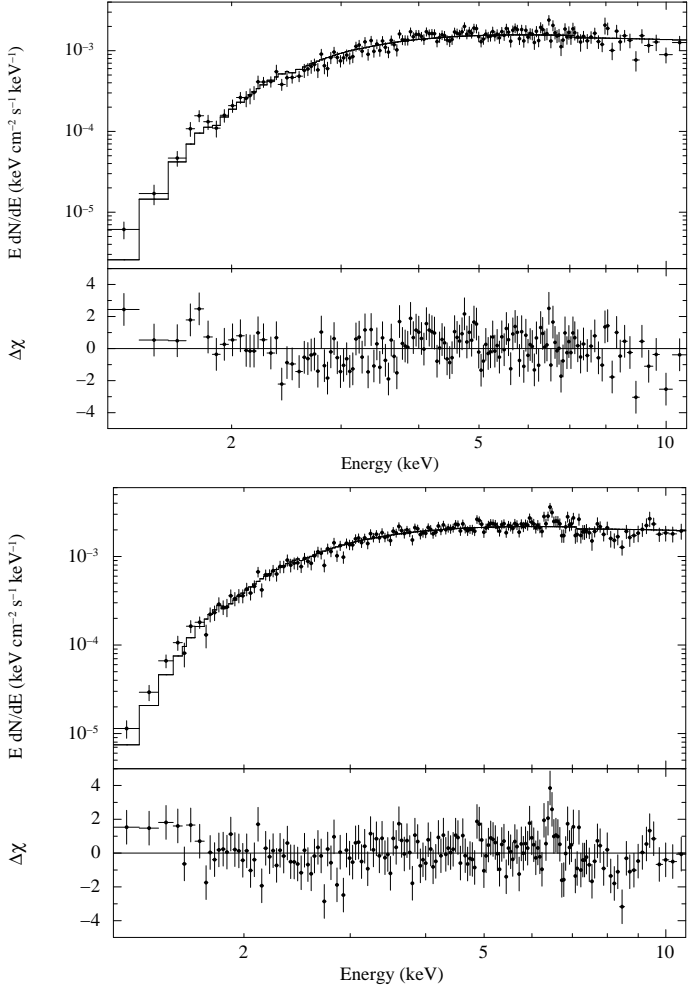

  \resizebox{\hsize}{!}{\includegraphics[angle=-90]{spectrum_401_euf.ps}}
  \resizebox{\hsize}{!}{\includegraphics[angle=-90]{spectrum_501_euf.ps}}
  \caption{Averaged \xmm\ Epic-pn spectra of \ax\ extracted from the observations  
    0084100401 (top) and 0084100501 (bottom). 
    In both cases we also show the best-fit model 
    (an absorbed power-law) and the residuals from these fits. In the observation 
    0084100501 the residuals clearly showed an iron line 
    at $\sim$6.4~keV.}  
  \label{fig:xmmspectrum2}
\end{figure}

In order to search for spectral variations with the source intensity, we extracted 
from observation 0084100401 two \xmm\ spectra by selecting time intervals 
in which the source count-rate in the 0.3-12 keV energy band was $<$2 and $>$2 cts/s. 
The first spectrum (effective exposure time 12~ks) could be reasonably well described 
($\chi^2_{\rm red}$/d.o.f.=1.2/125) by using an absorbed 
power-law (PL) model with $N_{\rm H}$=(6.3$^{+0.6}_{-0.5}$)$\times$10$^{22}$~cm$^{-2}$ 
and $\Gamma$=1.4$^{+0.2}_{-0.1}$ (uncertainties are given at 90\% c.l. throughout the paper). The estimated 1-10~keV X-ray flux was 
1.3$\times$10$^{-11}$~erg/cm$^2$/s. Alternatively, this spectrum could be 
well described ($\chi^2_{\rm red}$/d.o.f.=1.0/125) by using a 
blackbody (BB) model with a temperature of $kT_{\rm BB}$=(1.78$\pm$0.08)~keV and a radius 
of $R_{\rm BB}$=0.42$\pm$0.03~km (for an assumed distance of 10~kpc). 
The spectrum extracted at higher count-rates (effective exposure time 
2~ks) could be reasonably well fitted by using the same models discussed above. 
This gave $N_{\rm H}$=(6.2$^{+0.9}_{-0.8}$)$\times$10$^{22}$~cm$^{-2}$, 
and $\Gamma$=1.3$\pm$0.2 for the PL model and 
$N_{\rm H}$=(3.2$\pm$0.5)$\times$10$^{22}$~cm$^{-2}$, 
$kT_{\rm BB}$=(2.0$\pm$0.2)~keV, and $R_{\rm BB}$=0.60$\pm$0.07~km 
for the BB model. In this case the estimated flux is 
3.8$\times$10$^{-11}$~erg/cm$^2$/s (1-10~keV).  
This analysis did not reveal any significant change 
in the spectral parameters between the higher and lower count-rate 
spectra. We thus also extracted the average spectrum of this observation 
(see Fig.~\ref{fig:xmmspectrum2}).  
This could be well fitted with an absorbed PL model, the parameters of the fit 
are reported in Table~\ref{tab:log}. 
A similar good fit ($\chi^2_{\rm red}$/d.o.f.=0.9/147) could also be obtained by using a BB model  
($kT_{\rm BB}$=(1.85$\pm$0.07)~keV, $R_{\rm BB}$=0.45$\pm$0.03~km, 
$N_{\rm H}$=(3.3$\pm$0.3)$\times$10$^{22}$~cm$^{-2}$). 

We carried out a similar analysis for the observation 0084100501. 
The rate-resolved spectra were extracted during the time intervals in which 
the source count-rates were $>$2 and $<$2~cts/s (0.5-12~keV).   
%Beside an absorbed PL or BB continuum, 
%these spectra clearly showed the presence of an iron line at $6.4$~keV, whose equivalent 
%width marginally increased for lower count rates.  
We fitted both spectra with an absorbed power-law model and noticed that the residuals 
from these fits indicated the presence of an iron line at $\sim$6.4~keV (see Fig.~\ref{fig:xmmspectrum2}). 
We therefore added a Gaussian line to the spectral model used for the fit.  
The spectrum extracted at higher count rates (effective exposure time 4.4~ks) 
gave ($\chi^2_{\rm red}$/d.o.f.=1.1/134) a power-law photon index of $\Gamma$=1.3$\pm$0.1, an absorption column density of 
$N_{\rm H}$=(5.3$\pm$0.4)$\times$10$^{22}$~cm$^{22}$, an energy for the iron line 
centroid of $E_{\rm line}$=6.45$\pm$0.06~keV and an equivalent width  
of $EW$=0.06$_{-0.05}^{+0.04}$~keV (errors on the EWs are given at 68\% c.l. 
throughout the paper). The normalization of the line was (3.8$\pm$2.4)$\times$10$^{-5}$, 
 only yielding an indication of the presence of the line. 
For this spectrum the estimated 1-10~keV X-ray flux was 
(4.3$_{-0.3}^{+0.1}$)$\times$10$^{-11}$~erg/cm$^{2}$/s.  
By using a BB model instead of a PL, we obtained
$N_{\rm H}$=(2.7$\pm$0.3)$\times$10$^{22}$~cm$^{-2}$, $kT_{\rm BB}$=(1.95$\pm$0.08)~keV, and  $R_{\rm BB}$=0.63$\pm$0.04~km. 

The spectrum extracted at a lower count rate (effective exposure time 10.3 ks) gave 
$\Gamma$=1.4$\pm$0.1, $N_{\rm H}$=(4.6$\pm$0.4)$\times$10$^{22}$~cm$^{-2}$, 
$E_{\rm line}$=6.42$\pm$0.03~keV, $EW$=0.14$\pm$0.06~keV ($\chi^2_{\rm red}$/d.o.f.=1.0/132), and a flux of 
(1.7$_{-0.3}^{+0.1}$)$\times$10$^{-11}$~erg/cm$^{2}$/s (1-10 keV). The 
normalization of the line in this case was (3.3$\pm$1.1)$\times$10$^{-5}$, 
thus indicating a detection significance $>$3$\sigma$. 
By using an absorbed BB instead of the PL component to fit the 
lower count-rate spectrum in the observation 0084100501 would give 
$N_{\rm H}$=(2.2$\pm$0.2)$\times$10$^{22}$~cm$^{-2}$, $kT_{\rm BB}$=1.77$\pm$0.08~keV, and 
$R_{\rm BB}$=0.44$\pm$0.03~km. We also extracted the source 
spectrum by using the total exposure time of the observation 0084100501. 
This spectrum is shown in Fig.~\ref{fig:xmmspectrum2} and the best-fit parameters obtained 
with an absorbed PL model plus a Gaussian line are reported in Table~\ref{tab:log}. 
In this case we found $E_{\rm line}=6.44 \pm 0.03$~keV and an $EW$=0.09$\pm$0.03. 
The normalization of the line was (3.4$\pm$1.0)$\times$10$^{-5}$. 
By using a BB component instead of the PL for this spectrum would give 
$N_{\rm H}$=(2.4$\pm$0.2)$\times$10$^{22}$~cm$^{-2}$, $kT_{\rm BB}$=1.85$\pm$0.06~keV, 
$R_{\rm BB}$=0.51$\pm$0.03~km, $\chi^2_{\rm red}$=1.0/168 and parameters for the iron line 
fully in agreement (within the errors) with those reported above.   

\begin{figure}
  \resizebox{\hsize}{!}{\includegraphics[angle=-90]{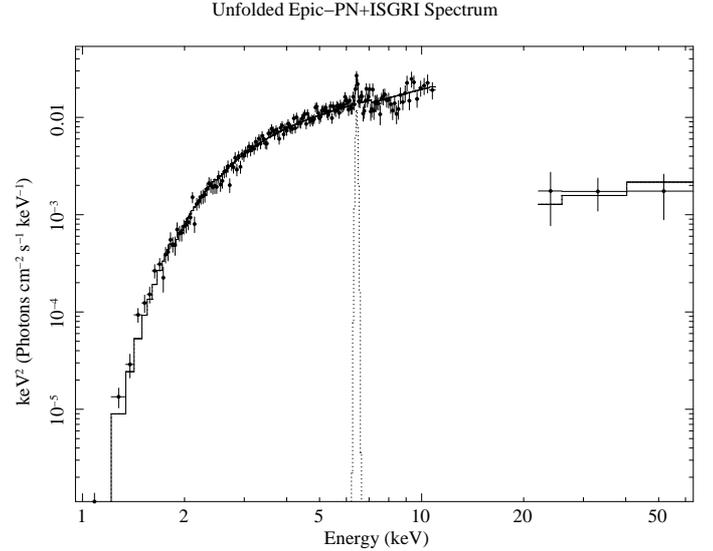}} 
  \caption{Unfolded Epic-pn+ISGRI spectrum of \ax.\ The best-fit model is obtained 
    with an absorbed power-law model plus a Gaussian line at $\sim$6.4~keV (see text for 
    details).}
  \label{fig:unfolded}
\end{figure}

In order to compare the results found for the two \xmm\ observations of \ax,\ we added an iron line 
with a centroid energy fixed at 6.44~keV to the spectrum extracted by using the total exposure time 
of observation 0084100401 and determined the 90\% c.l. upper limit on its normalization at 1.5$\times$10$^{-5}$. 
This value is lower than the one measured in observation
0084100501, but still compatible with that expected due to the lower flux ($\sim$40\%)
of the source in the observation 0084100401. We thus conclude that it
 is not possible to infer unambiguously from the present data a variation of the iron line 
parameters between the two observations.
Even though no simultaneous \xmm\ and \inte\ observation were available, we tried a fit 
to the combined averaged ISGRI spectrum and the Epic-pn averaged spectrum of the observation ID.~0084100501. 
This spectrum could be well described with an absorbed power law model, and we introduced 
a normalization constant to take into account both the intercalibration between the Epic-pn and 
ISGRI instruments and the variability of the source. The values of the absorption column density and power-law photon index were found to 
be fully consistent with those of the averaged Epic-pn spectrum, but the normalization 
constant turned out to be 0.04$\pm$0.02 (this value would be 0.06$\pm$0.03 if the same fit were performed by using the \xmm\ spectrum of the observation 0084100401). 
The relatively low value of the normalization 
constant between the two instruments indicates that on average the X-ray flux 
of the source is much lower than that measured during the \xmm\ observations. This is consistent 
with the results reported in Table~\ref{tab:log}. The unfolded Epic-pn+ISGRI spectrum is shown 
in Fig.~\ref{fig:unfolded}.  

Finally, we also extracted for both \xmm\ observations a spatially resolved X-ray 
spectrum by selecting different regions centered on the source in order to search for 
spectral shape variations and investigate a possible extended nature of the source. 
We did not find any evidence for a significant change in the spectral 
parameters, and concluded that the elongated shape of the source is most likely caused by
distortion of the instrument PSF for sources close to the border of the FOV (see also above). 
Because of this problem, the determination of an improved source position requires some caution. 
We used the automatic \xmm\ source-detection task 
{\sc edetect-chain} (with manually fine-tuned parameters), to have an estimate of the source position and error 
in the two \xmm\ observations for both the Epic-pn and Epic-MOS2 cameras. In the observation 
0084100401, the two best-determined positions are 
(i) $\alpha_{\rm J2000}$=19$^{\rm h}$10$^{\rm m}$43$\fs$42, 
$\delta_{\rm J2000}$=09${\degr}$16$\arcmin$29$\farcs$3   
and 
(ii) $\alpha_{\rm J2000}$=19$^{\rm h}$10$^{\rm m}$43$\fs$37, 
$\delta_{\rm J2000}$=09${\degr}$16$\arcmin$29$\farcs$3, 
for the Epic-pn and the Epic-MOS2, respectively. 
The corresponding positions obtained in the observation 0084100501 were (i) 
$\alpha_{\rm J2000}$=19$^{\rm h}$10$^{\rm m}$43$\fs$44, 
$\delta_{\rm J2000}$=09${\degr}$16$\arcmin$30$\farcs$7 and 
(ii) $\alpha_{\rm J2000}$=19$^{\rm h}$10$^{\rm m}$43$\fs$39, 
$\delta_{\rm J2000}$=09${\degr}$16$\arcmin$30$\farcs$0. 
%The error returned by the {\sc edetect-chain} task is of only $\sim$0.2'' 
%in all cases, but this is only a statistical error and does not take into account the uncertainties 
%in the pointing of the spacecraft. 
According to the latest calibration document available, the relative astrometry among all EPIC cameras is 
estimated to be better than 1.5'', and the absolute astrometric accuracy of any source in the \xmm\ FOV is of 
$\sim$2''. Only for faint MOS sources near the detection limit this error can be as large as 4'' 
(90\% c.l.). In the present case all results on the 
positions we obtained are consistent within 1.5'' as expected. Given the position of \ax\ at the very rim of the 
\xmm\ detector, we assume in the following as the best determined position of the source: 
$\alpha_{\rm J2000}$=19$^{\rm h}$10$^{\rm m}$43$\fs$39,  
$\delta_{\rm J2000}$=09${\degr}$16$\arcmin$30$\farcs$0 (J2000), with a conservative 
associated error of 2'' (90\% c.l.).  
In Sect.~\ref{sec:counterparts} we use this result to search 
for possible counterparts to \ax.\ 

A Fourier analysis of the \xmm\ data of \ax\ did not reveal any
 indication for a possible coherent periodicity.

\subsubsection{ASCA}
\label{sec:asca}  

\ax\ was serendipitously observed in several \asca\ \citep{tanaka94} observations performed 
in 1993 and 1999. A log of these observations is provided in Table~\ref{tab:log}. 
We used data from the two gas-imaging spectrometers \citep[GIS2 and GIS3,][]{ohashi96, makishima96} 
and applied standard screening criteria\footnote{See the ASCA ABC guide at 
http://heasarc.gsfc.nasa.gov/docs/ asca/abc/abc.html.}. 
We extracted the source light curves and spectra from a circular 
region of 2' radius. A larger extraction region could not be used because of the emission from the nearby SNR 
(see also Sect.~\ref{sec:xmm}). For the background, we used a similar extraction region, located in  
the same part of the FOV as the source events, as described in the \asca\ ABC guide. 
Note that \ax\ is located well within the Galactic disk and thus we could not use the \asca\ blank-sky observations 
(see pag.~80 of the \asca\ ABC guide). We used the latest available GIS2 and GIS3 instrument response 
files (gis2v4\_0.rmf and gis3v4\_0.rmf) and generated for each observation the corresponding ancillary file 
with the tool {\sc ascaarf}. 

Only in observation 50005000 was the statistics 
sufficiently high to rebin the spectrum to have at least 20 photons per bin and perform a 
minimum $\chi^2$ fitting. In all observations where the source was detected but the statistics was 
relatively poor, the C-statistics was used for the fits. We also determined a 90\% c.l. upper limit 
on the source X-ray flux in all the observations in which \ax\ was not detected.  In these cases, we obtained  
a source and background spectrum by using the same extraction regions adopted in the other observations, and 
fitted them with an absorbed PL model. The absorption column density and PL photon index were fixed to those of  
observation 50005000, and the 90\% c.l. error on the model normalization was used to obtain an upper limit on the 
flux (1-10 keV energy band). All these results are reported in Table~\ref{tab:log}. 
Given the low S/N of the \asca\ observations of \ax,\ we did not investigate possible timing 
features in these data.

\subsubsection{Chandra}
\label{sec:chandra}  

A research in the {\sc heasarc} data archive\footnote{http://heasarc.gsfc.nasa.gov/.} revealed 
that \ax\ was also twice observed by the ACIS telescope on-board \chan\ \citep{garmire03}. 
The first of these observations (ID.~117) was performed on 2001 July 8 and lasted for 54~ks. 
\ax\ was serendipitously observed in the ACIS-I2 chip, but not detected. Unfortunately, 
the analysis of this observation revealed the presence of a large number of poorly illuminated columns 
on the ACIS-I2 chip (most likely a read-out problem), and therefore we did not consider this 
observation in our subsequent analysis. 
In the second observation (ID.~9615), performed on 2008 May 31 and lasting 1.65~ks, 
\ax\ was observed in the FOV of the ACIS-S3 chip, but not detected. 

We derived from these data an upper limit on the source X-ray flux by using both the {\sc aprates} 
task\footnote{See the on-line thread: http://cxc.harvard.edu/ciao/threads/aprates/$\!$ index.html\#netcts.} 
available within the {\sc ciao} package (v.4.2) and by building a fluxed image of the ACIS-S3 
chip\footnote{See the on-line thread: http://cxc.harvard.edu/ciao/threads/eff2evt.}.  
The first method allowed us to derive a 68\% c.l. upper limit on the source count rate 
of 0.012 cts/s (0.5-7~keV), which corresponds to a 1-10 keV X-ray flux of 
4.0$\times$10$^{-13}$~erg/cm$^{2}$/s (we used the online tool 
{\sc webpimms}\footnote{http://heasarc.nasa.gov/Tools/w3pimms.html} and assumed a PL model with $\Gamma$=1.4 and 
$N_{\rm H}$=4.8$\times$10$^{22}$~cm$^{-2}$ to be consistent with the other upper limits 
derived, see Sect.~\ref{sec:asca}). 
The second method permits us to directly calculate the energy flux in units of erg/cm$^2$/s for each 
event in the selected \chan\ chip, taking into account the quantum efficiency and effective area as well. 
With this method we estimated a 3$\sigma$ upper limit on the source X-ray flux of 
5.4$\times$10$^{-13}$~erg/cm$^2$/s (0.5-7~keV) energy band, compatible with the limit derived with 
the first method.

\subsubsection {Counterparts of \ax\ }
\label{sec:counterparts}   

\begin{figure}
  \resizebox{\hsize}{!}{\includegraphics{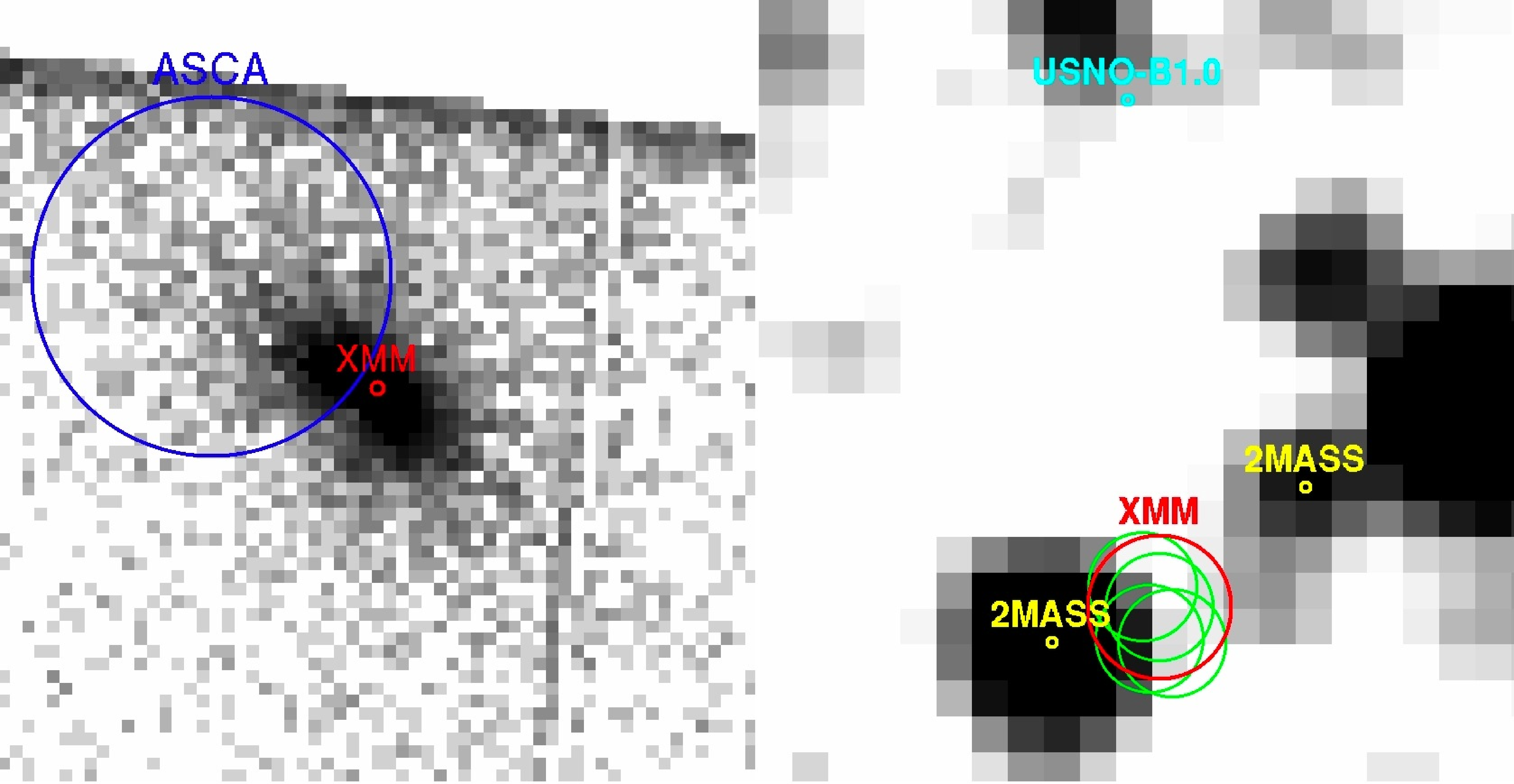}}
  \caption{\xmm\ (left) and infrared (2MASS, J band, right) image of the region around 
    \ax.\ We show the error circles (green) associated with the determined 
    positions in the 2 \xmm\ observations for the Epic-pn and Epic-MOS2 cameras, and the best 
    adopted source position (red circle, error 2'' at 90\% c.l.). We also show in blue  
    the previously determined \asca\ position (only slightly outside the \xmm\ error circle) and the closest 
    2MASS and USNO-B1.0 counterparts discussed in Sect.~\ref{sec:counterparts} (see the electronic version of 
    the paper for the colored picture).}
  \label{fig:axj19counterpart}
\end{figure}

We used the improved source position found from the \xmm\ data to search 
for an infrared and/or optical counterpart to \ax.\ In Fig.~\ref{fig:axj19counterpart} we show 
the \xmm\ FOV around \ax,\ and the corresponding 2MASS image \citep{2mass}. The four determined positions 
in the \xmm\ observations (green circles) and the best adopted source position 
(red circle, error 2'' at 90\% c.l.) are also shown. 
We found only one likely counterpart,  2MASS\,J19104360+0916291, which partly lies within the 
\xmm\ error circle and has J=15.044$\pm$0.030, H=13.000$\pm$0.022, and K=11.808$\pm$0.023. No cataloged optical 
counterparts were found within the \xmm\ error circle. The closest object in the USNO-B1.0  
catalog is shown in the figure but because of the relatively large separation from \ax,\ it 
is very unlikely that it is associated with \ax.\ We note that the lack of a clear optical counterpart 
is compatible with the high 
absorption in the direction of the source measured from the \xmm\ observations (see Sect.~\ref{sec:xmm}). 
We also queried the FIRST Survey and the NVSS catalogs in search for a radio counterpart, 
but did not find any obvious candidate.  

\begin{figure}[h!]
\hspace*{0.7cm}
    \includegraphics[width=0.45\textwidth]{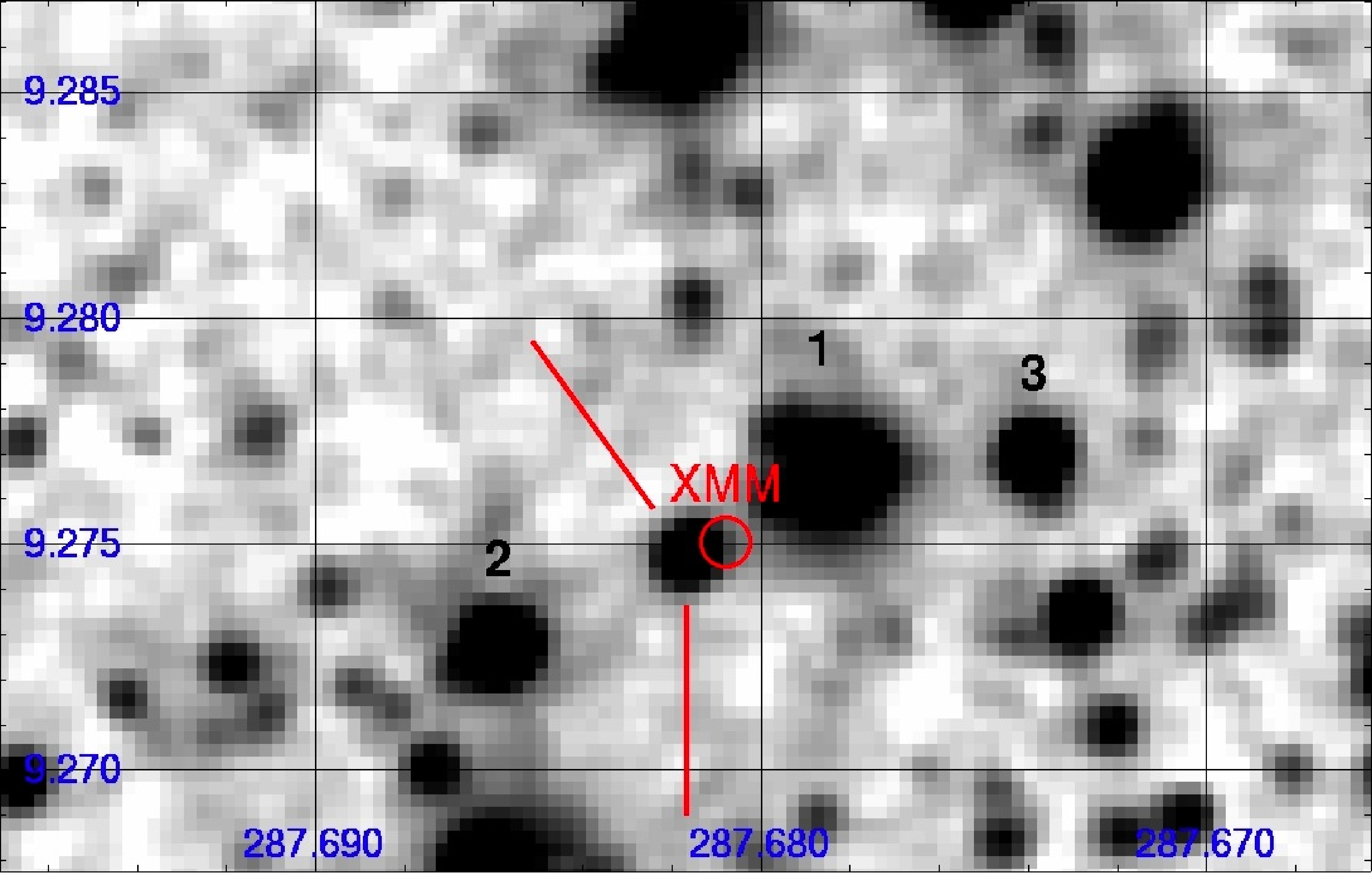}
    \resizebox{\hsize}{!}{\includegraphics{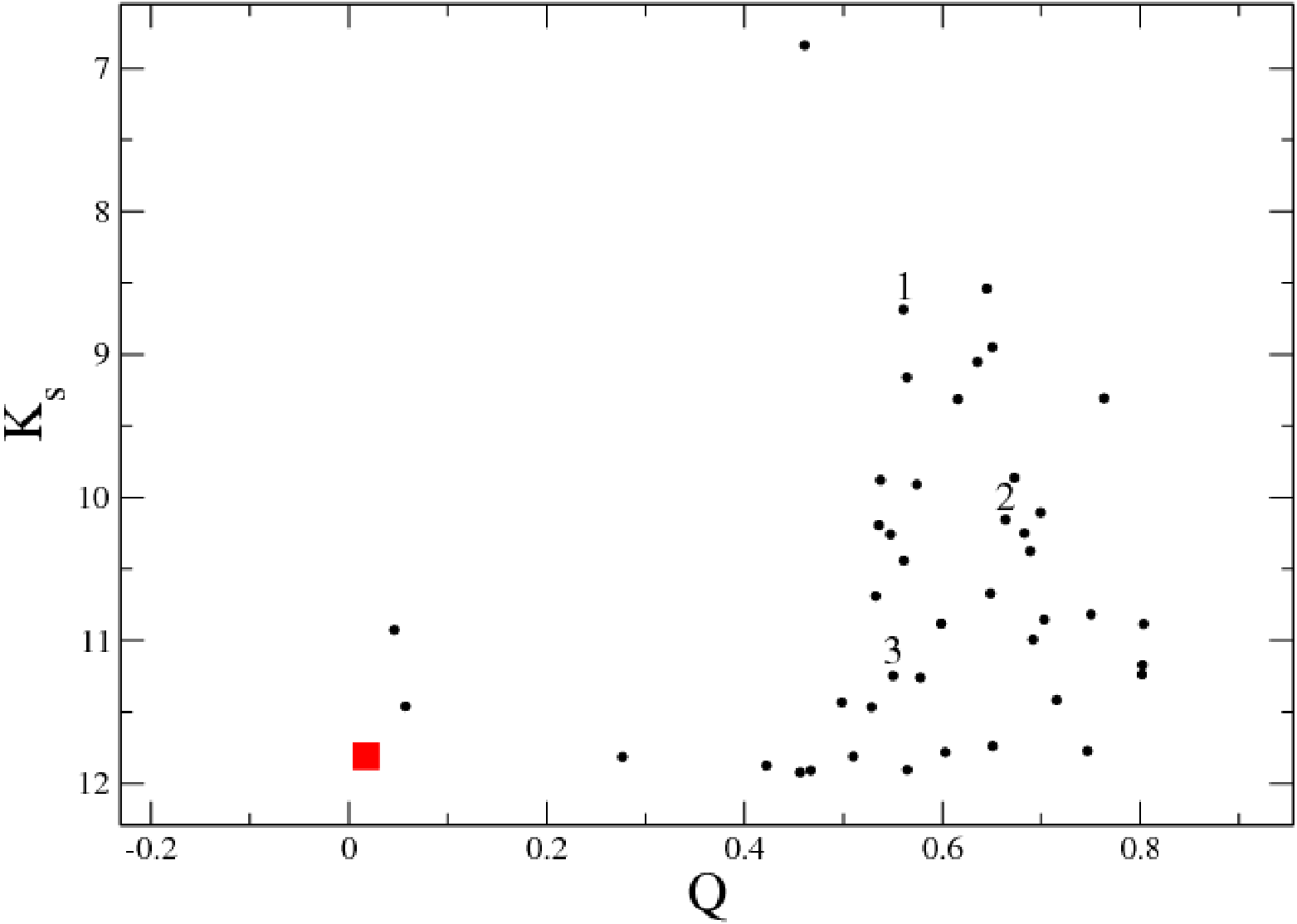}}
    \caption{{\it Top:} 2MASS image (K band, coordinates in Right Ascension and Declination) of the FOV around \ax.\ 
      The best-determined 
      \xmm\ position is marked in red, and a few nearby potential IR candidates are also displayed. 
      {\it Bottom:} Plot of $Q$ vs. $K_{s}$ magnitudes for the 2MASS stars around the
      XMM position (see text for details). The square represents the likely IR counterpart 
      to \ax.\ Some of the nearby potential candidates are also shown.}  
  \label{fig:2mass} 
\end{figure}

To investigate the nature of the possible IR counterpart found above for 
\ax\ in more detail, we applied the analysis described in \citet{negueruela07} to the objects in the 
FOV of 2MASS\,J19104360+0916291. For each star represented in Fig.~\ref{fig:2mass} we used 2MASS photometric data  
to obtain the values of their J, H, and K. Then, we calculated the quantity 
$Q$=($J$-$H$)-1.70($H$-$K_s$) for each object (here $K_s$=K) and plotted these values as a function of 
$K_s$ in the bottom panel of Fig.~\ref{fig:2mass}. According to \citet{negueruela07}, the objects 
characterized by $K_{s}$$\lesssim$11 and $Q$$\lesssim$0.2 are promising supergiant candidates. 
Even though this method does not provide a secure classification, it suggests that the nature of the 
source 2MASS\,J19104360+0916291 (red square in Fig.~\ref{fig:2mass}) 
would be compatible with being a supergiant star. 
We discuss this aspect further in Sect.~\ref{sec:discussion}.  

\section{New \inte\ sources} 
\label{sec:newsources} 

In the IBIS/ISGRI FOV around \ax,\ we found three new sources that had previously remained  
undetected. These appeared to be the only excesses 
found independently both in the OSA9.0 mosaic and the mosaic obtained with the {\sc bat\_imager} software 
\citep{segreto10,segreto10b}. %[][private communication from A.~Segreto]
%One of these newly found sources is coincident with the unclassified \einst\ source 2E\,1912.5+1031 \citep{hertz88}.
 A summary of the properties of the sources is given in Table~\ref{tab:newsources}.
There we report the best values of the detection significance,  
source position, and the associated errors.
The sources were named according to the standard \inte\ convention. 
A mosaic containing all the new sources is shown in Fig.~\ref{fig:mosaic}.

\begin{table}[h]
  \caption{\label{tab:newsources}Newly discovered \inte\ sources around \ax.\ } 
  \centering
 \scriptsize
  \begin{tabular}{@{}ccccccc@{}}
    \hline\hline
\noalign{\smallskip}
Name & RA  & Dec & Err. & Det.\tablefootmark{a} & Counts\tablefootmark{b}  & Exp.\tablefootmark{c} \\    
      & (deg) & (deg) & (')  & ($\sigma$)  & & (Ms) \\ 
\noalign{\smallskip}
\hline
\noalign{\smallskip}
IGR\,J19173+0747  & 289.349 & 7.785 &  2.1 & 10.0 & 0.15$\pm$0.02  & 3.0 \\
\noalign{\smallskip}
IGR\,J19294+1327 & 292.367 & 13.459 & 3.4 & 6.8 & 0.12$\pm$0.02 & 2.1 \\
\noalign{\smallskip}
IGR\,J19149+1036 & 288.73\tablefootmark{d}& 10.61\tablefootmark{d}&  1.0\tablefootmark{d}& $\sim$20\tablefootmark{d} & $\sim$0.3\tablefootmark{d}&   2.6\\
\hline 
\end{tabular}
\tablefoot{
\tablefoottext{a}{Detection significance in the IBIS/ISGRI mosaic (17-80~keV);}
\tablefoottext{b}{The count rates in cts/s estimated from the ISGRI mosaic. In this energy band 1~mCrab=0.28 cts/s.}
\tablefoottext{c}{In this energy band 1~mCrab=0.28 cts/s.}
\tablefoottext{d}{These values are affected by large systematic uncertainties related to the presence of GRS 1915-105.}
}
\end{table}

We report in Table~\ref{tab:newsources} only a first-order approximation for the values 
of IGR\,J19149+1036 because it is relatively close ($\lesssim$20~arcmin) 
to the brighter object GRS\,1915-105 and a precise determination of the degree 
of contamination would require a much more detailed analysis. 
We note, though, that the inferred source position  is coincident with the \einst\ source 2E\,1912.5+1031. 
For IGR\,J19173+0747 and IGR\,J19294+1327, we performed a spectral analysis. 
For IGR\,J19173+0747, the ISGRI spectrum was well described  
by a power-law model with $\Gamma$=3.3$_{-0.7}^{+0.9}$ ($\chi^2_{\rm red}$/d.o.f.=0.3/5) 
and flux of $F_{\rm 20-40~keV}$=5.6$\times$10$^{-12}$~erg/cm$^2$/s. 
A similar analysis for IGR\,J19294+1327 gave $\Gamma$=2.6$_{-0.7}^{+0.8}$, 
$F_{\rm 20-40~keV}$=6.5$\times$10$^{-12}$~erg/cm$^2$/s and 
$\chi^2_{\rm red}$/d.o.f.=0.4/4.

\subsection{\swift\,/XRT follow-up of the newly discovered sources}

Following the discovery of the three new sources reported in the previous section, 
we asked for follow-up observations (PI L. Stella) in the soft X-ray domain with 
\swift\,/XRT \citep[0.3-10~keV;][]{gehrels04}. At the time of writing, only the follow-up observations 
in the direction of IGR\,J19173+0747 and IGR\,J19294+1327 were carried out. 
In order to search for the X-ray counterpart of these two sources, we used the best-determined 
positions and errors in Table~\ref{tab:newsources}.
We processed all the \swift\,/XRT data with 
the {\sc xrtpipeline} and the latest calibration files available (caldb v. 20091130). 
All observations were performed in photon-counting mode (PC).  
Filtering and screening criteria were applied by using {\sc ftools} (Heasoft v.6.9). 
We extracted source and background light curves and spectra by selecting event 
grades of 0-12 for the PC mode, and created the exposure maps for each observation 
through the {\sc xrtexpomap} task. We rebinned each spectrum where possible to have 
at least 20 photons per bin and used the latest spectral redistribution matrices available 
in the {\sc heasarc} calibration database (v.011). 
Ancillary response files, accounting for different 
extraction regions, vignetting and PSF corrections, were generated 
with  the {\sc xrtmkarf} task. For each of the two sources, 
we determined an improved source position by using the {\sc xrtcentroid} task. 
\begin{figure}
  \includegraphics[width=0.5\textwidth]{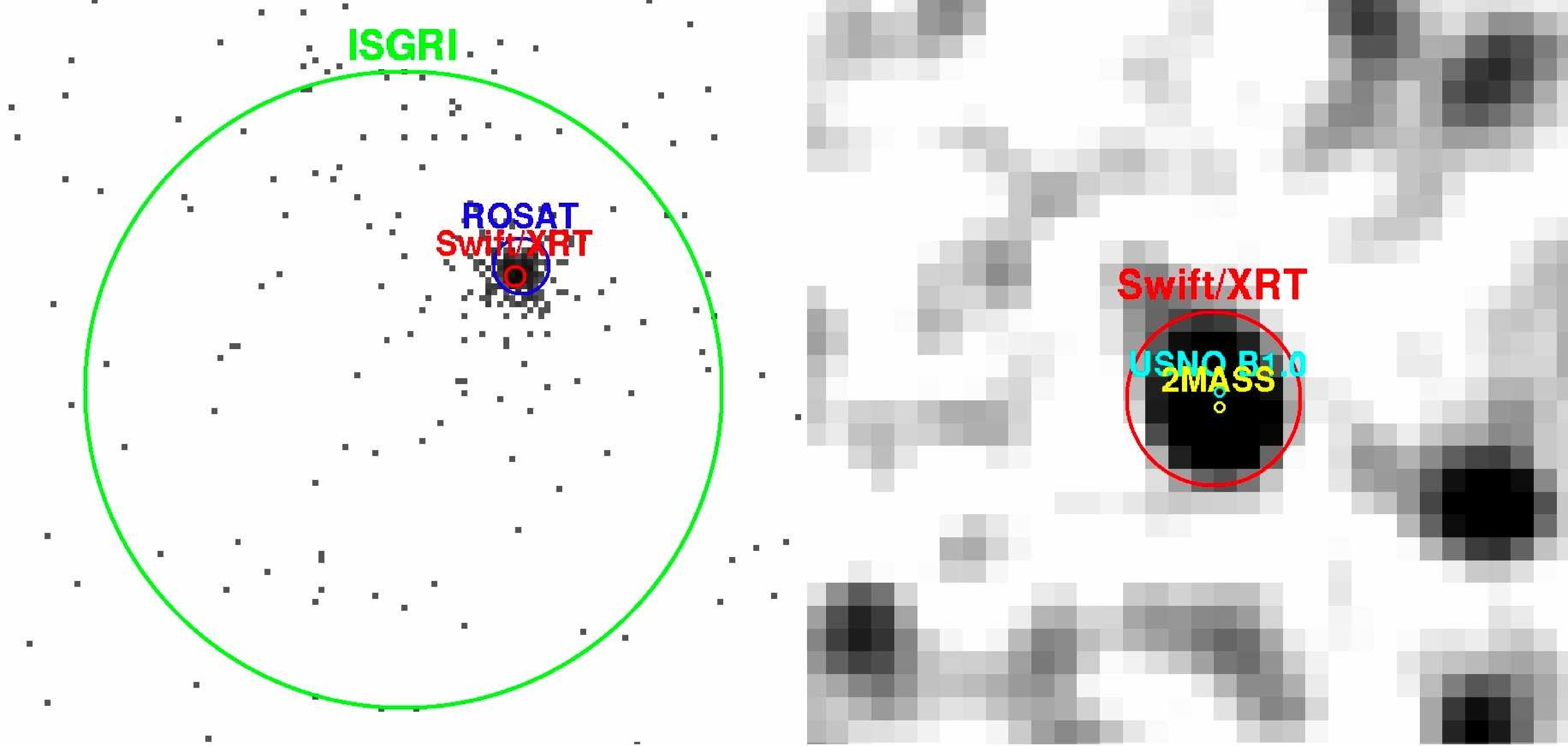}
  \includegraphics[width=0.5\textwidth]{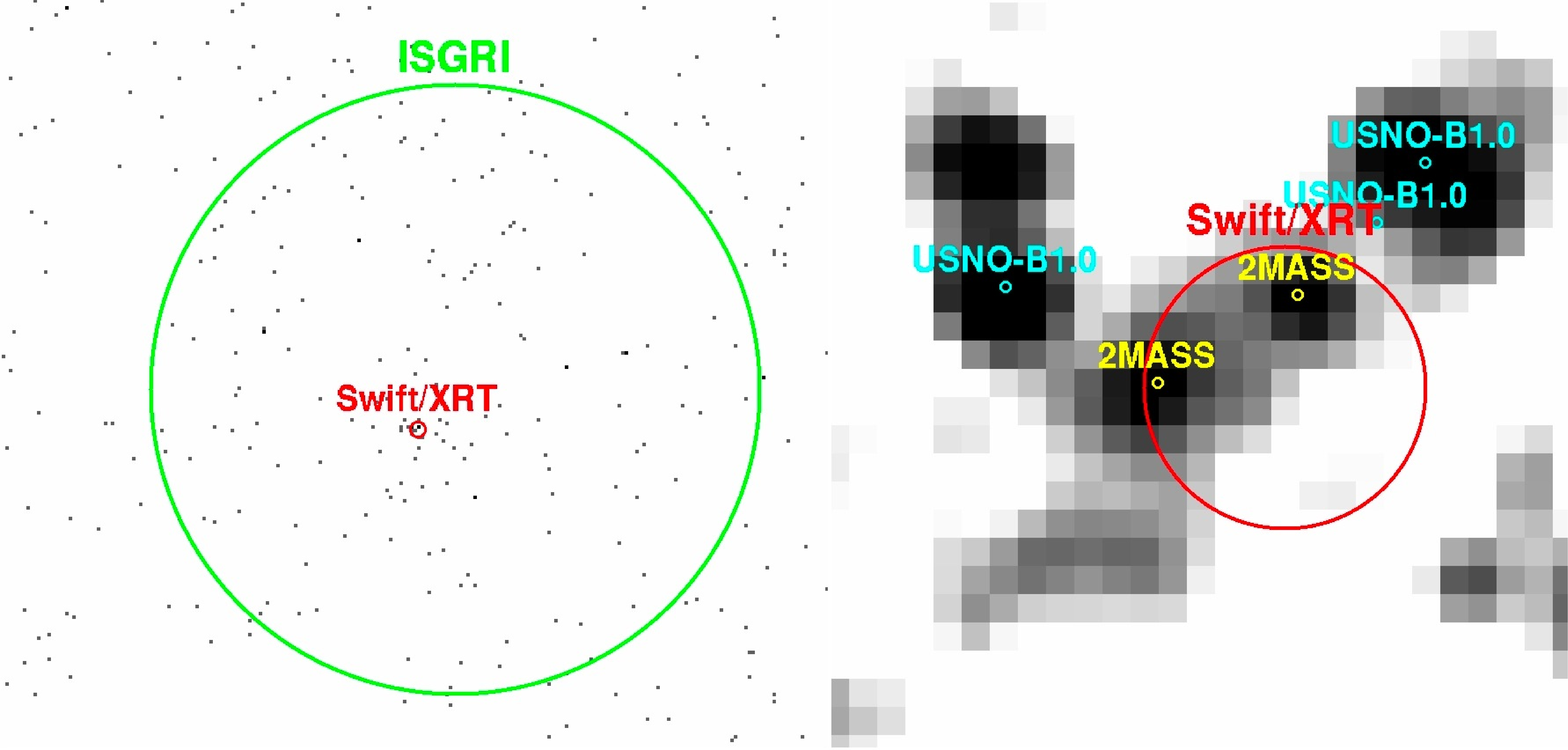}
  \caption{\swift\,/XRT observations of the newly discovered \inte\ sources. {\it Top}: 
  IGR\,J19173+0747. In this case only one soft X-ray source is found in \swift\,/XRT within the 
  ISGRI error circle. We show on the right the 2MASS infrared image (J band) together with the \swift\,/XRT 
  error circle and the position of the IR and optical counterpart. A positionally coincident \rosat\ 
  source is also shown.  
  {\it Bottom}: For IGR\,J19294+1327 only one very faint soft X-ray excess has been found in the \swift\,/XRT 
  observation (S/N=3.7). Within the assumed XRT error circle there are two possible IR candidates, but no optical counterpart has been 
  found (see the electronic version of the paper for the colored image).}  
  \label{fig:swift}
\end{figure}

\subsubsection{IGR\,J19173+0747}
\label{sec:new3} 

IGR\,J19173+0747 was observed by \swift\,/XRT starting on 2010 February 22 at 
08:07:00, for a total exposure time of 6~ks. An image of the source is shown in 
Fig.~\ref{fig:swift}. Inside the \inte\ error circle there is only one soft X-ray 
source. The spectrum of the source could be well fitted by an absorbed power-law 
model. We estimated a PL photon index of $\Gamma$=0.6$\pm$0.2 and obtained an upper limit on the 
absorption column density of $N_{\rm H}$$<$6$\times$10$^{21}$~cm$^{-2}$ (90\% c.l.). 
The corresponding flux is (6$_{-1.5}^{+1.0}$)$\times$10$^{-12}$~erg/cm$^2$/s (0.5-10~keV).
Extrapolating this flux to the 20-40~keV band would predict a much higher flux ($\sim 2.8 \times 10^{-11}$~erg/cm$^2$/s)
than the one observed with IBIS/ISGRI (5.6$\times$10$^{-12}$~erg/cm$^2$/s, see Sect.~\ref{sec:newsources}). 
This, together with the different photon index derived in the two energy ranges,
suggests a break in the spectrum at energies between 10 and 20~keV, or alternatively, 
a variability of the source.
We obtained a refined source position at $\alpha_{\rm J2000}$=19$^{\rm h}$17$^{\rm m}$20$\fs$8 and 
$\delta_{\rm J2000}$=07${\degr}$47$\arcmin$51$\farcs$1, with an associated uncertainty of 
3.8~arcsec (90\% c.l.). This position is consistent with that of the \rosat\ source 
1RXS\,J191720.6+074755 \citep{voges99}. 
The count rate of the source is 0.04$\pm$0.01 cts/s in the 0.1-2.4~keV band.
By assuming the spectral shape determined above, we checked with the online tool {\sc webpimms} that
the \rosat\ countrate would correspond to a flux compatible with that of \swift\,/XRT.
Inside the \swift\,/XRT error circle we found only one possible NIR and optical counterpart. 
The NIR counterpart is 2MASS\,J19172078+0747506, characterized by 
J=13.945$\pm$0.031, H=13.520$\pm$0.030, and K=13.311$\pm$0.043. The optical counterpart 
is USNO-B1.0\,0977-0532587 (R1=15.46, B1=16.91, R2=14.99, B2=16.14, I=14.78). 
We queried the FIRST Survey and the NVSS catalogs in search for a radio counterpart, 
but did not find any obvious candidate.

\subsubsection{IGR\,J19294+1327}
\label{sec:new6}

IGR\,J19294+1327 was observed by \swift\,/XRT twice, on 2010 February 22 beginning at 23:59:01 
and on 2010 February 26 beginning at 10:16:01. The total exposure time was 7.4~ks. In the \swift\,/XRT 
FOV only one very faint X-ray source is visible within the 
ISGRI error circle (S/N=3.7, see Fig.~\ref{fig:swift}). 
Given the relatively low S/N ratio, other observations are needed to confirm this detection.
Assuming that this is the real counterpart of the \inte\ source, we estimated its best position
at $\alpha_{\rm J2000}$=19$^{\rm h}$29$^{\rm m}$29$\fs$80 and 
$\delta_{\rm J2000}$=13${\degr}$27$\arcmin$05$\farcs$44 (associated error of 
5.0~arcsec). The estimated source count rate was (4.0$\pm$1.1)$\times$10$^{-3}$~cts/s, corresponding to 
an X-ray flux of 3$\times$10$^{-13}$~erg/cm$^2$/s (0.3-10 keV, we assumed $N_{\rm H}$=7.1$\times$10$^{21}$~cm$^{-2}$, and $\Gamma$=1.5). 
This low flux, compared to that measured in the 20-40~keV band, suggests that either the source is intrinsically
very faint in the 0.3-10 keV range, or that it is a strongly variable source.
Within the \swift\,/XRT error circle we found the two possible NIR counterparts 2MASS\,J19292976+1327087 
(J=16.075$\pm$0.143,	H=15.187$\pm$0.150, K=14.761$\pm$0.133) and 2MASS\,J19293011+1327056 
(J=16.018$\pm$0.137,	H=14.725$\pm$0.119, K=13.803$\pm$0.101). No cataloged optical or radio 
counterpart was found at these positions.

\section{Discussion and conclusions}
\label{sec:discussion}

We reported on the analysis of all available X-ray observations carried 
out with \inte,\ \xmm,\ \chan,\ and \asca\ that included \ax\ in the instruments' FOVs. 
The information we could extract from these data showed that in the soft X-ray energy band (0.5-10~keV) 
the source is clearly variable. The highest dynamic range in the X-ray flux we could investigate with all  
available observations and instruments is $\gtrsim$60 (see Table~\ref{tab:log}).  
The highest flux was recorded during the \xmm\ observations, which also provided an improved source position 
and the first detailed characterization of the spectrum of the source. 
In the \xmm\ data, \ax\ appeared to be variable on a relatively short timescale (hundreds of seconds) 
and the 0.5-10~keV X-ray spectrum could be well fitted with an absorbed power-law 
model. In the second available \xmm\ observation, we also found that an iron line was required 
to fit the spectrum of the source. The centroid of the line is at $\sim$6.4~keV, consistent with  
fluorescence origin from cold iron (likely due to iron material in a ionization state not higher that Fe XX).  
The absorption column density measured in the different X-ray observations 
showed only minor changes and remained always much higher than that expected in the direction of the source 
($\sim$1.5$\times$10$^{22}$~cm$^{-2}$). The power-law photon index was consistent with being constant 
in all data we analyzed. In the hard X-ray energy band (17-80~keV) the source was characterized  
by a mean X-ray flux of a few 10$^{-12}$~erg/cm$^2$/s. 
No evidence was found in the \xmm\ and \inte\ data for a coherent periodicity that could be associated 
with the spin period of a neutron star hosted in this system or an orbital period. 
The improved X-ray position obtained thanks to the \xmm\ observations also allowed us to search for possible counterparts in 
different energy bands (optical, infrared, radio). In Sect.~\ref{sec:counterparts} we suggested a possible association 
between \ax\ and the object 2MASS\,J19104360+0916291, which is the closest classified object to the \xmm\ 
position. No obvious counterpart in the optical and radio band could be identified. 

The available X-ray data on \ax\ do not allow for an unambiguous classification of this source. 
Its position, relatively close to the Galactic plane, favors the hypothesis of a Galactic source  
and we discuss below a few different possibilities.\  
%LMXB
A relevant feature, for investigating the nature of \ax\ is the iron line 
visible in the \xmm\ observations. The width and the centroid of the line 
are compatible with a fluorescence origin and thus suggest that \ax\ is likely part 
of a binary systems. Similar iron lines are indeed unlikely to appear in the X-ray spectra 
of isolated compact objects and magnetars \citep[see, e.g.][]{mereghetti2008}. 
The apparent lack of intrinsic broadening 
would also argue against \ax\ being a low-mass X-ray binary \citep[][and references therein]{bhat07}.
Among the different subclasses of cataclysmic variables \citep[for a review see][and references therein]{warner95}, 
polar systems are hardly detected in the ISGRI energy band. 
Furthermore, for these systems a variability in flux of order $\sim 60$ is uncommon. 
Similar arguments apply to the intermediate polar case. These sources sometimes display  
emission above 20~keV, but they are known to be generally persistent objects in the soft energy band (0.5-10~keV). 
We also suggest that the X-ray properties of \ax\ are not compatible with those of the novae sources 
\citep{warner95}. 
%classical HMXB
A more likely possibility is that \ax\ is a new member of the high-mass X-ray binaries (HMXB) discovered 
by \inte.\ According to this interpretation, the iron line observed in the \xmm\ spectrum could originate from  
irradiation of cold iron in the wind of a massive companion. 
The analysis conducted in Sect.~\ref{sec:counterparts} on the nature of the possible IR counterpart to 
\ax\ and the low Galactic latitude of the source would also support this interpretation.  
However, the association of \ax\ with the class of the HMXBs would still face some difficulties. 
Among the different subclasses of HMXBs observed by \inte,\ neither the classical supergiants 
\citep[see e.g.,][]{walter06} nor the SFXT \citep{sguera05} have a behavior fully compatible 
with that of \ax.\ 
On the one hand, the variability in the X-ray flux reported in Table~\ref{tab:log} for the different 
observations available clearly shows that \ax\ is not a persistent source as expected for the 
classical supergiant HMXBs. On the other hand, the non detection in the ISGRI data of some  
bright and short flares typical of the SFXT sources argues against this interpretation.  
%Be X-ray binaries
Another possibility is that \ax\ is a Be X-ray binary, and the \xmm\ observation luckily 
caught the source during an outburst. According to this interpretation we would expect 
a somewhat higher flux during this period in the \inte\ observations of \ax.\  
We checked that no simultaneous \inte\ and \xmm\ observations were available, and in 
the closest ISGRI data (obtained about 7 days before and after the \xmm\ observations) 
\ax\ was detected with a count rate compatible with the average one. 
We concluded that if the high flux measured in the \xmm\ data corresponded to an outburst, 
then this should have lasted less than 10~days. This could still be compatible 
with the durations of the outbursts observed from Be systems. Extrapolating the \xmm\ spectrum in the ISGRI energy band (17-80 keV), we estimated that a similar bright event should correspond to an ISGRI count rate of 1.5-2~cts/s. 
We searched for similar events in all available ISGRI data by using a 5-day binned lightcurve and found no evidence for such bursts.
During the 2200 days spanned by the ISGRI data the source was effectively monitored for approximately 415 days. This suggests 
that \ax\ should spend $<80$\% of the time in the bright X-ray state observed by \xmm.\ 
Additional pointed observations in X-rays with \xmm\ and \chan,\ 
as well as follow-up observations in different  
energy bands, are required in order to firmly establish the nature of \ax.\  

Besides carrying out a detailed study of \ax\ in X-rays, we also report the discovery of 
three new hard X-ray sources in the IBIS/ISGRI FOV around \ax.\  
These sources were independently detected with the OSAv9.0 and the 
{\sc bat\_imager} software (A.~Segreto, private communication) and are the only excesses 
appearing in both the mosaics extracted around \ax\ with the two software packages.
As we showed in Sect.~\ref{sec:newsources}, a detailed study of  
2E\,1912.5+1031 could not be carried out because of the 
likely contamination in the \inte\ data by the bright source GRS\,1915-105 and 
the lack of proper follow-up with \swift.\  A more refined analysis of this source will be 
reported elsewhere. 
For the two new \inte\ sources IGR\,J19173+0747 and IGR\,J19294+1327, we identified a counterpart 
in the soft X-ray energy band (0.3-10~keV) thanks to dedicated 
\swift\ observations (see Sect.~\ref{sec:newsources}), and searched for possible cataloged optical and infrared 
counterparts. However, owing to the relative faintness of the sources in the hard X-ray band 
($<$10$^{11}$~erg/cm$^2$/s in the 17-80~keV band) and the short exposure time of the \swift\,/XRT observations, 
a clear classification of the two sources is still premature, and other observations are required 
in order to determine their nature.

\begin{acknowledgements} 
We thank G. Cusumano for sharing the information 
on the \swift\,/BAT data on \ax,\ A.Segreto for the information on the 
results obtained with {\sc bat\_imager} software, and M. Falanga for useful discussions. 
We thank the referee P.R. den Hartog for his useful comments that improved the content of this paper.
LS acknowledges financial support from ASI.  
\end{acknowledgements}

\bibliographystyle{aa}
\bibliography{pavan}

\end{document}